\documentclass[prd,twocolumn,aps,letterpaper,amsmath,amssymb,preprintnumbers,showpacs,superscriptaddress,floatfix,longbibliography,nofootinbib]{revtex4-2}
\usepackage{array}
\usepackage{graphicx}
\usepackage[export]{adjustbox}
\usepackage[caption=false]{subfig}
\usepackage{tikz}
\usetikzlibrary{tikzmark}
\usetikzlibrary{calc}
\allowdisplaybreaks % let equations split over multiple pages
\usepackage[hypertexnames=true]{hyperref}
\usepackage[capitalize]{cleveref}
\hypersetup{
    colorlinks=true,       % false: boxed links; true: colored links
    linkcolor=blue,          % color of internal links
    citecolor=blue,        % color of links to bibliography
    filecolor=blue,      % color of file links
    urlcolor=blue           % color of external links
}

\usepackage{slashed}
\usepackage{mathtools}
\usepackage{listings}
\usepackage{pgf}
\usepackage{fancyvrb}
\usepackage{longtable}
\usepackage{bm}
\usepackage{braket}
\usepackage{tikz-cd} 
\usepackage{amsthm}
\usepackage{mathrsfs}
\usepackage{algorithm}
\usepackage{algpseudocode}

\newtheorem{theorem}{Theorem}
\newtheorem{lemma}{Lemma}

\DeclareMathOperator{\imag}{Im}

\DeclareMathOperator{\discontinuity}{disc}

\newcommand{\R}[0]{\ensuremath{\mathbb{R}}}
\newcommand{\C}[0]{\ensuremath{\mathbb{C}}}

\newcommand{\ball}[0]{\ensuremath{\mathbb{B}}}
\newcommand{\disk}[0]{\ensuremath{\mathbb{D}}}

\newcommand{\vecl}[0]{\bm{a}} % Used in statement and proof of Lemma 1
\newcommand{\vecr}[0]{\bm{b}} % Used in statement and proof of Lemma 1
\newcommand{\vacuum}[0]{\Omega} % Used in 
\newcommand{\hankel}[0]{H}

\newcommand{\abs}[1]{\ensuremath{| #1|}}
\newcommand{\norm}[1]{\left\lVert#1\right\rVert}

\renewcommand{\Im}{\ensuremath{\mathrm{Im}}}

\begin{document}
\title{Moment problems
and bounds for matrix-valued smeared spectral functions}
\begin{abstract}
Numerical analytic continuation arises frequently in lattice field theory, particularly in spectroscopy problems.
This work shows the equivalence of common spectroscopic problems to certain classes of moment problems that have been studied thoroughly in the mathematical literature.
Mathematical results due to Kovalishina enable rigorous bounds on smeared matrix-valued spectral functions, which are implemented numerically for the first time.
The required input is a positive-definite matrix of Euclidean-time correlation functions; such matrices are routinely computed in variational spectrum studies using lattice quantum chromodynamics.
This work connects the moment-problem perspective to recent developments using the Rayleigh--Ritz method and Lanczos algorithm.
Possible limitations due to finite numerical precision are discussed.
\end{abstract}

\author{Ryan Abbott}
\email{rabbott@mit.edu}
\affiliation{Center for Theoretical Physics--A Leinweber Institute, Massachusetts Institute of Technology, Cambridge, MA 02139, USA}
\author{William~I.~Jay}
\email{william.jay@colostate.edu}
\affiliation{Department of Physics, Colorado State University, Fort Collins, CO 80523, USA}
\author{Patrick~R.~Oare}
\email{poare@bnl.gov}
\affiliation{Physics Department, Brookhaven National Laboratory, Upton, NY 11973, USA}
\preprint{MIT-CTP/5896}

\maketitle

\section{Introduction}

The problem of numerical analytic continuation from a finite set of points arises frequently in lattice field theory.
One example, and the focus of the present work, is reconstructing smeared spectral functions from Euclidean-time correlation functions.
Another example appears in the finite-volume formalism for hadron spectroscopy, where one seeks to constrain scattering amplitudes in the complex energy plane given information at real energies~\cite{Briceno:2017max, Salg:2025now}.

The starting point for the present work is the K{\"a}ll{\'e}n--Lehmann spectral representation of a Green's function $G(s)$ appearing in quantum field theory (QFT)~\cite{Kallen:1952zz,Lehmann:1954xi,Schwartz:2014sze}, 
\begin{align}
    G(s) = \int_0^\infty ds' \frac{\rho(s')}{s'-s + i\epsilon}, \label{eq:KL-spectral-rep}
\end{align}
where $\rho$ is the spectral function satisfying
\begin{align}
    \rho(s) = \frac{1}{\pi}\Im\,G(s) = \frac{1}{2\pi i} \discontinuity G(s). \label{eq:stieltjes-perron}
\end{align}
\Cref{eq:KL-spectral-rep} makes manifest that evaluating the Green function at complex $z=\omega+i\epsilon$ is equivalent to computing a smeared spectral function $\rho_\epsilon(\omega)$ using a Poisson kernel
\begin{align}
    \delta_\epsilon(\omega)  = \frac{1}{\pi}\frac{\epsilon}{\omega^2-\epsilon^2} = \frac{1}{\pi}\Im \frac{-1}{\omega + i \epsilon},
\end{align}
a property emphasized in Refs.~\cite{Bergamaschi:2023xzx,Poggio:1975af}.
Smeared spectral functions have appealing physical interpretations and are often accessible experimentally~\cite{Davier:2005xq,Davier:2019can} and through numerical calculation using lattice field theory~\cite{Hansen:2017mnd,Bulava:2021fre}.
On the other hand, from a mathematical perspective, evaluating \cref{eq:KL-spectral-rep} for $z\in\C$ is nothing more than the \emph{Stieltjes transform} 
\begin{align}
    G(z) = \int_{\R} dx\,\frac{\rho(x)}{x-z}. \label{eq:stieltjes-trans-def}
\end{align}
The Stieltjes transform clearly satisfies
\begin{align}
    G(z)=G^\dagger(z^*), &&
    \frac{G(z)-G^\dagger(z)}{z-z^*} \succeq 0 \label{eq:stieltjes_positvity}
\end{align}
for $z \in \C \setminus \R$ and the notation $M \succeq 0$ indicates that the matrix $M$ is positive-definite.
This transform arises naturally in the study of continued fractions, interpolation theory, and asymptotic series (see Ref.~\cite{KJELDSEN199319} for a historical overview).
The present work focuses on the connection of the Stieltjes transform to moment problems, which are defined in \cref{sec:moment_problems} below.

Connections between the Stieltjes transform and Pad{\'e} approximations have been discussed previously in the context of lattice field theory~\cite{Peris:2006ds,Aubin:2012me}.
For instance, Ref.~\cite{Aubin:2012me} used the Stieltjes transform to express the hadronic vacuum polarization using a convergent sequence of Pad{\'e} approximants. 

Similar ideas have been explored recently in the condensed-matter literature.
Ref.~\cite{nogaki2023nevanlinnajl} has discussed the Stieltjes transform and the connection to the Hamburger moment problem, focusing on a frequency-space conception of the problem.
Ref.~\cite{Fei_2021} has considered analytic continuation of matrix-valued Green functions in frequency space.
In contrast to Refs.~\cite{nogaki2023nevanlinnajl,Fei_2021}, the present work connects moments of matrix-valued Green functions directly to the Euclidean-time correlation functions.

The main results of the present work can be summarized as follows.
First, the generic problem of determining a matrix-valued spectral function from a positive-definite matrix of Euclidean-time correlation functions can be viewed as a Hamburger moment problem.
Second, existing results in the mathematical literature due to Kovalishina can be used to characterize the space of possible solutions to a given moment problem~\cite{Kovalishina1984}.
To the authors' knowledge, the present work gives the first numerical realization of Kovalishina's analytic results.
Third, these methods provide rigorous bounds on smeared matrix-valued spectral functions.
Finally, the moment-problem perspective helps connects recent developments in the physics literature on spectral reconstructions, including the Rayleigh--Ritz method, the Lanczos algorithm, and Nevanlinna--Pick interpolation.

The remainder of the work has the following structure.
\Cref{sec:moment_problems} provides a basic statement of moment problems. 
\Cref{sec:lqcd} makes the connection between moments problems and positive-definite matrices of Euclidean-time correlation functions calculable in lattice field theory.
\Cref{sec:mathematics} discusses Kovalishina's general solution of the truncated Hamburger moment problem in terms of the Stieltjes transform~\cite{Kovalishina1984}, including rigorous bounds for individual components of the smeared matrix-valued spectral function.
\Cref{sec:other_methods} provides some connections to recent ideas regarding the Rayleigh--Ritz method and the Lanczos algorithm.
\Cref{sec:numerical_results} presents several numerical examples.
\Cref{sec:noise} discusses some practical considerations regarding input data with finite numerical precision.
\Cref{sec:conclusions} closes with some conclusions.
\Cref{sec:technical} collects various technical results.

\section{Moment Problems \label{sec:moment_problems}}

The definition of a moment problem begins with the notion of a positive distribution, i.e., a matrix-valued function $\rho_{ab}(x) \succeq 0$ on the real line with indices $a,b \in \{0, \dots, N-1\}$, where $N$ denotes the size of the matrix.\footnote{
The mathematical literature on moment problems typically writes $d\mu(x) \equiv \rho(x)\, dx$ and phrases theorems in the language of measure theory; that level of abstraction will not be required in the present work.
}
The inequality is understood to mean that for each $x\in\R$, the matrix $\rho_{ab}(x)$ is positive semidefinite.
The scalar case corresponds to a familiar positive distribution $\rho(x) \geq 0$ on the real line.
The indices $a,b$ will frequently be suppressed when no confusion will arise.
The $n$th moment with respect to the distribution is defined via
\begin{align}
    C_n = \int dx\, x^n \rho(x) \label{eq:moment-def},
\end{align}
where $n$ is a non-negative integer.
Provided the integrals converge, no conceptual complications arise in the ``forward problem" of computing the moments given a distribution.

The inverse problem---inferring an unknown distribution $\rho(x)$ given the moments $C_n$---is known as a \emph{moment problem}.
Moment problems of particular interest for the present work are the Hausdorff moment problem ($\rho$ with support on the bounded interval $[0,1]$), the Stieltjes moment problem ($\rho$ with support on the positive real line), and the Hamburger moment problem ($\rho$ with support anywhere on the real line).
In practice, the focus of attention will be on the Hamburger moment problem, treating the others as special cases.

A moment problem is referred to as \emph{truncated} if only a finite subset of the moments is given. 
Since any numerical calculation is by necessity finite, all moment problems in the present work are understood to be truncated unless otherwise stated.
Textbook treatments of moment problems include Refs.~\cite{akhiezer2020classical,schmudgen2017moment}.

The connection between moment problems and the  Stieltjes transform $G(z)$ is well known~\cite{akhiezer2020classical,schmudgen2017moment} and can be seen from the asymptotic expansion
\begin{align}
    G(z) = -\sum_{n=0}^\infty \frac{C_n}{z^{n+1}} \label{eq:stieltjes_asymptotic}
\end{align}
which follows immediately from Taylor expansion of \cref{eq:stieltjes-trans-def} around $z=\infty$.
In other words, the Stieltjes transform acts as a generating function for the moments of the distribution $\rho$.
\cref{eq:stieltjes-perron}, known in the mathematical literature as the Stieltjes--Perron inversion formula, then states that the Stieltjes transform encodes the same information as the original distribution.
Given this equivalence, an alternative characterization of a moment problem is to recover $G(z)$ from asymptotic data specified by the moments $C_n$.

The asymptotic expansion in \cref{eq:stieltjes_asymptotic} is typically made precise in the mathematical literature by imposing the limiting condition
\begin{align}
\lim_{z\to i\infty} z^{n+1} \left( G(z) + \sum_{0\leq m\leq n-1} \frac{C_m}{z^{m+1}}\right) = -C_n. \label{eq:hamburger_interpolation_condition}
\end{align}
Any choice of positive distribution leads to a Stieltjes transform $G(z)$ satisfying \cref{eq:hamburger_interpolation_condition}.
H.~Hamburger has shown that the converse holds as well, namely, that a function $G(z)$ satisfying the asymptotic condition \cref{eq:hamburger_interpolation_condition} along with \cref{eq:stieltjes_positvity} can be represented via \cref{eq:stieltjes-trans-def} for some positive distribution~\cite{Hamburger:1920}.
This classic result can also be found in textbooks, e.g., Theorem 3.2.1 of Ref.~\cite{akhiezer2020classical}.
The importance of \cref{eq:hamburger_interpolation_condition} is that it allows the moment problem to formulated as an interpolation problem with nodes at $\pm i\infty$.

\begin{table*}[t!]
    \centering
    \caption{Summary of moment problems arising in lattice field theory.
    The moment problems are written in terms of the density $\tilde{\rho}_{ab}(\lambda)$ defined in \cref{eq:density_lambda}, where $\lambda$ is the real variable in the complex transfer-matrix-eigenvalue plane.
    The density is related to the physical spectral function via 
    \cref{eq:physical_spectral_function}.
    \label{tab:moment_problem_summary}
    }
    \begin{tabular}{c c c c}
    \hline\hline
        Physical Problem & Spectral Decomposition & Moment problem & Domain for $\lambda$\\
        \hline
        Zero-temperature ($\beta=\infty)$ spectroscopy & \cref{eq:laplace_transform} & Hausdorff & $[0, 1]$\\
        Finite-temperature spectroscopy & \cref{eq:finite_temp_exp} & Stieltjes & $[0,\infty]$\\
        Finite-temperature staggered spectroscopy & \cref{eq:staggered} & Hamburger & $[-\infty, \infty]$\\
    \hline\hline
    \end{tabular}
\end{table*}

\section{Moment Problems in Lattice Field Theory\label{sec:lqcd}}

Consider a lattice QFT defined in a finite spatial volume with Euclidean temporal extent $\beta$ and lattice spacing $a=1$.
The transfer matrix~\cite{Luscher:1976ms,Luscher:1984is} is defined in terms of the Hamiltonian $\hat{H}$,
\begin{align}
    \hat{T} = e^{-\hat{H}},
\end{align}
and evolves the system by one unit in Euclidean time.
Numerical applications of lattice field theory typically
enjoy direct access neither to the infinite-dimensional transfer matrix nor to the Hamiltonian. 
Instead, knowledge of the theory is expressed in terms of positive-definite $N \times N$ matrices of gauge-invariant correlation functions 
\begin{align}
    C_{ab}(t) = \langle\hat{\mathcal{O}}_a(t)\hat{\mathcal{O}}^\dag_b(0)\rangle
    = \langle \psi_a | \hat{T}^t |\psi_b\rangle,
    \label{eq:corr-func-def}
\end{align}
where $a,b \in \{0, \dots, N-1\}$ are integers that index the operators and $\ket{\psi_a} = \mathcal{O}_a^\dagger \ket{\vacuum}$ represents the state excited by acting with the operator $\hat{\mathcal{O}}_a^\dagger$ on the vacuum $\ket{\vacuum}$.
Precise forms for the operators will not be needed for the present discussion.

Let $\ket{k}$ be the eigenvectors of $\hat{T}$ with associated eigenvalues $\lambda_k \in (0, 1]$ labeled by $k \in \mathbb{N}$, so that
\begin{equation}
   \hat{T} \ket{k} = \lambda_k \ket{k}.
\end{equation}
The eigenvalues $\lambda_k$ of the transfer matrix are related to the energies $E_k$ by 
\begin{equation}
\label{eq:lambda-E-relation}
\lambda_k = e^{-E_k}.
\end{equation}
The eigenvectors can be taken to be orthonormal, satisfying the normalization condition
\begin{equation}
    \braket{k | k'} = \delta_{kk'}.
\end{equation}
The initial vectors $\ket{\psi_a}$ can be expanded in terms of the eigenvectors,
\begin{equation}
\label{eq:psi-spectral-expand}
\ket{\psi_a} = \sum_{k=0}^\infty Z_{ka} \ket{k},
\end{equation}
where $Z_{ka} = \braket{k | \psi_a}$. Substituting \cref{eq:psi-spectral-expand} into \cref{eq:corr-func-def} and using orthogonality gives the correlation function in the form~\cite{Wagman:2024rid}
\begin{equation}
    \begin{aligned}
    C_{ab}(t) &= \sum_{k,k'} 
    Z_{ka}^* Z_{kb}
    \braket{k | T^t | k'} \\
    &= \sum_k Z_{ka}^* Z_{kb} \lambda_k^t.
    \end{aligned}
\end{equation}
The correlation function can then be expressed as
\begin{equation}
    C_{ab}(t) = \int_0^1 d \lambda \, \lambda^t \tilde{\rho}_{ab}(\lambda) \label{eq:correlator_moment_problem}
\end{equation}
with a density $\tilde{\rho}_{ab}(\lambda)$ defined by
\begin{equation}
    \tilde{\rho}_{ab}(\lambda) \equiv \sum_k Z_{ka}^* Z_{kb}
    \delta(\lambda - \lambda_k). \label{eq:density_lambda}
\end{equation}
\Cref{eq:correlator_moment_problem} is exactly the expression for the $t$-th moment of $\tilde{\rho}_{ab}$.
Therefore the Hausdorff moment problem associated with the matrix of Euclidean-time correlation functions $C_{ab}(t)$ coincides exactly with the spectroscopic problem of determining the values of $Z_{ka}$ and $\lambda_k$ (see \cref{tab:moment_problem_summary}).

The density $\tilde{\rho}_{ab}(\lambda)$ is closely related to the familiar spectral function $\rho_{ab}(\omega)$ defined at zero temperature by
\begin{align}
    C_{ab}(t) &= \int_0^\infty d\omega \, \rho_{ab}(\omega) e^{-\omega t} \label{eq:laplace_transform}\\
    \rho_{ab}(\omega) &= \sum_k Z_{ka}^* Z_{kb} \delta(\omega - E_k).
\end{align}
This definition implies that
\begin{align}
\rho_{ab}(\omega) = \lambda(\omega)\,\tilde{\rho}_{ab}(\lambda(\omega)), \label{eq:physical_spectral_function}
\end{align}
a result which follows either from the familiar composition rule for the Dirac delta function, $\delta(\lambda(\omega) - \lambda_k) = \delta(\omega - E_k)/\lambda_k$, or equivalently from enforcing consistency between \cref{eq:correlator_moment_problem} and \cref{eq:laplace_transform}
\begin{align}
    \int_0^1 d\lambda \, \tilde{\rho}_{ab}(\lambda)
    &= \int_0^\infty d\omega \, \rho_{ab}(\omega) = 
    \sum_k Z_{ka}^* Z_{kb},
\end{align}
with Jacobian $d\omega = d\lambda/\lambda$. In the continuum limit ($a \to 0$), a fixed physical energy $\omega$ will approach zero in lattice units so that $\lambda(\omega) = e^{-\omega} \to 1$, which implies that the Jacobian factor is trivial in the continuum limit.

\begin{figure*}[t!]
    \centering
    \includegraphics[width=\linewidth]{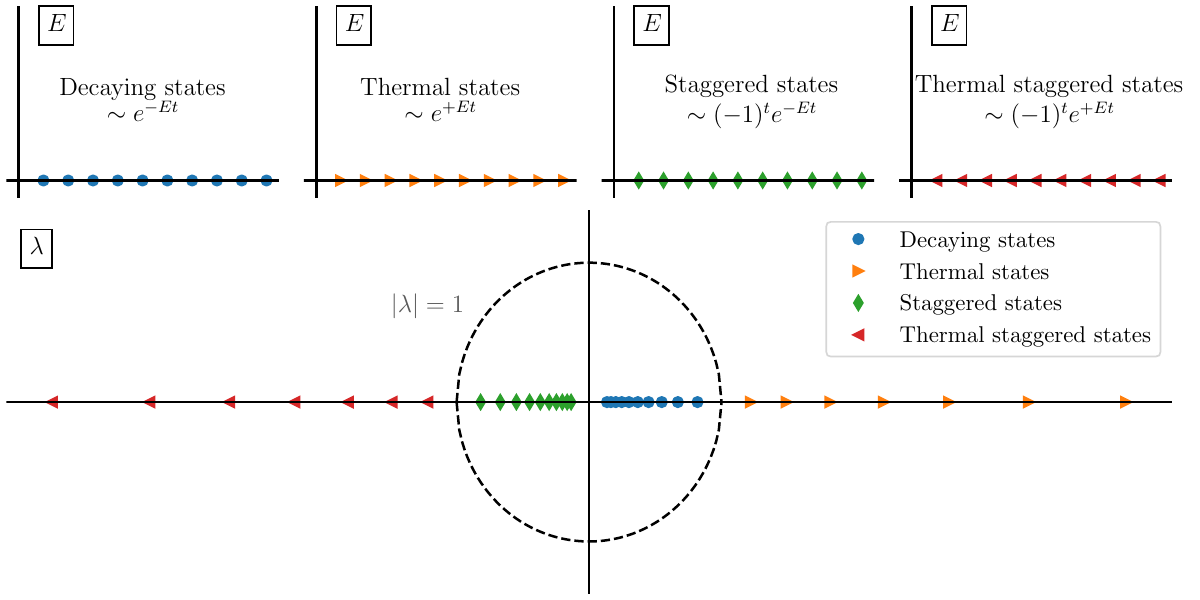}
    \caption{
    The mapping of energy eigenvalues (upper row) to the complex transfer-matrix-eigenvalue plane (lower row).
    Standard decaying and thermal eigenvalues are mapped to $\lambda >0$.
    Staggered states are mapped to $\lambda <0$.
    }
    \label{fig:eigenvalue_plane}
\end{figure*}

Similar considerations apply to finite-temperature correlation functions, e.g., of the familiar form
\begin{align}
C_{ab}(t) &= \int_{0}^\infty d\omega\, \rho_{ab}(\omega) \frac{\cosh \omega (\beta/2 -t)}{\sinh \beta\omega/2}\\
&=  \int_{0}^\infty d\omega\, \rho_{ab}(\omega) 
\left( 
\frac{e^{-\omega t} + e^{-\omega(\beta-t)}}{1 - e^{-\beta\omega}}
\right).\label{eq:finite_temp_exp}
\end{align}
One can split the integrand into two terms scaling as $\lambda^t=e^{-\omega t}$ and
$(\lambda^\prime)^t=e^{+\omega t}$, both with $\omega >0$~\cite{Wagman:2024rid}.
Clearly $\lambda\in(0,1)$ and $\lambda' \in (1,\infty)$.
Combining the two terms into a single integrand over $(0,\infty)$ gives
\begin{align}
    C_{ab}(t) = \int_0^\infty d\lambda\, 
    \frac{\tilde{\rho}_{ab}(\lambda)}{|1-\lambda^\beta|} \lambda^t
\end{align}
This result shows explicitly that thermal correlation functions are moments of a positive distribution on the positive real-$\lambda$ line and therefore correspond to the Stieltjes moment problem.

Calculations in numerical lattice gauge theory frequently employ the staggered-fermion discretization, for which the zero-temperature spectral decomposition also includes ``oscillating" contributions of the form
\begin{align}
\begin{split}
    \sum_{k} Z_{ka}^* Z_{kb} (-1)^t e^{-E_k t}
    &=
    \int_0^\infty d\omega\, \rho_{ab}(\omega) \left(- e^{-\omega } \right)^t\\
    &= \int_{-1}^0 d\lambda\, \tilde{\rho}_{ab}(\lambda) \lambda^t,
\end{split}
\end{align}
where $\lambda = - e^{-\omega}$.
Repeating the preceding arguments for a thermal correlator 
\begin{align}
    C_{ab}(t) = \sum_k Z_{ka}^* Z_{kb} (-1)^{p_k t} \left( e^{-E_k t} + e^{-E_k(\beta- t)} \right) \label{eq:staggered},
\end{align}
where $p_k \in \{0, 1\}$ indicates whether the state $k$ is oscillating, shows that the generic thermal staggered-fermion spectroscopy problem can be identified with the Hamburger moment problem.

\Cref{tab:moment_problem_summary} summarizes the different moment problems associated with the spectral decompositions in \cref{eq:laplace_transform,eq:finite_temp_exp,eq:staggered}.
\cref{fig:eigenvalue_plane} summarizes the mapping of energy eigenvalues to the complex transfer-matrix-eigenvalue plane.
It bears emphasizing that expressions like \cref{eq:corr-func-def} involving $\hat{T}^t$ between external states are well known.
What seems not to have been mentioned explicitly in the literature is the connection between the generic (thermal, staggered) Euclidean-time correlation function, the Hamburger moment problem, and the physical spectral function in \cref{eq:physical_spectral_function}.
The remainder of the paper will focus on the Hamburger moment problem, treating the Hausdorff and Stieltjes problems as special cases in which the spectral function $\tilde{\rho}(\lambda)$ is defined on the full real-$\lambda$ line but happens to vanish on the appropriate intervals.

\section{Mathematical Solution of Moment Problems\label{sec:mathematics}}

A remarkable fact about certain interpolation problems in complex analysis is that a complete description of possible solutions can be given.
For instance, Nevanlinna--Pick interpolation is concerned with constructing an analytic function $f:\disk\to\disk$ such that $f(\zeta_n)=w_n$ at specified nodes $\zeta_n$ and values $w_n$, with $n\in\{1,2,\dots \}$.
Analyticity severely constrains the possible values of $f(\zeta)$ away from the interpolation nodes.
Nevanlinna~\cite{Nevanlinna1919,Nevanlinna1929} showed that, for each point $\zeta\in\disk$, $f(\zeta) \in \Delta_n(\zeta) \subset \disk$ where $\Delta_n(\zeta)$ is a disk known as the Wertevorrat.
Building on Ref.~\cite{PhysRevLett.126.056402}, Ref.~\cite{Bergamaschi:2023xzx} showed that the Wertevorrat can be used to provide rigorous uncertainty bounds for smeared spectral functions.
\Cref{sec:moment_problems} described how moment problems can also be formulated as interpolation problems with nodes at $\pm i \infty$. 
Consequently, moment problems exhibit similar constraints, as will be discussed below.

\subsection{Existence of solutions\label{sec:solution_existence}}

Solutions to a moment problem need not always exist.
For example, consider the scalar case and suppose that $C_0 =1$ so that $\rho$ is a probability distribution.
Positivity of the variance then implies that a valid set of moments must satisfy
\begin{align}
    C_0 C_2 - C_1^2 = \det \begin{bmatrix}
        C_0 & C_1 \\
        C_1 & C_2
    \end{bmatrix} \geq 0.
\end{align}

A similar positivity result holds quite generally~\cite{reed1975fourier}.
Let $\alpha_n \in \C$ be a set of arbitrary coefficients.
The inequality
\begin{equation}
    0 \leq \int dx\, \left|\sum_n \alpha_n x^n\right|^2 \rho(x)
    = \sum_{n,m} \alpha_n^* \alpha_m C_{n + m}
\end{equation}
implies the Hankel matrix
\begin{equation}
\hankel_{n,m} = C_{n + m}
    = \begin{bmatrix}
        C_0 & C_1 & C_2 & \dots \\
        C_1 & C_2 & C_3 & \dots \\
        C_2 & C_3 & C_4 & \dots \\
        \vdots & \vdots & \vdots & \ddots \\
    \end{bmatrix}_{n,m} \label{eq:hankel_matrix_definition}
\end{equation}
must necessarily be positive-semidefinite.\footnote{
The notation in \cref{eq:hankel_matrix_definition} is to be understood as saying that the element of $H$ in the $n$th row and $m$th column is the $(n+m)$th moment $C_{n+m}$.
In lattice field theory, $C_{n+m}$ will correspond to matrix of correlators at Euclidean time $n+m$, i.e. $C(n + m)$
}
This condition turns out to be sufficient for the scalar Hamburger moment problem.
The matrix version of the truncated Hamburger moment problem has a solution if and only if the corresponding block Hankel matrix is \emph{non-negatively extendable}, i.e., if and only if the block Hankel matrix can be enlarged to a block Hankel matrix of larger size which is still positive semi-definite~\cite{CHEN1998199}.
Positivity requirements on Hankel matrices like these are well known and have been recently noted in the physics literature, including Refs.~\cite{Abbott:2025yhm,Yu_2024,Chen:2004rq}.

Mathematical treatments of moment problems are also concerned with the question of determinacy.
In other words, for a given infinite (non-truncated) problem, ``Is the solution unique?"
An important result in Carleman's condition~\cite{akhiezer2020classical}, which roughly says that the problem will be determinate as long as the moments do not grow super-exponentially.
Since Euclidean-time correlators are bounded, moments problems arising in the continuum limit of lattice-field-theory calculations will presumably converge to a unique solution, although proof of this conjecture exceeds the scope of the present discussion.

\subsection{$J$-theory approach to interpolation problems}

Methods for solving interpolation problems in complex analysis, going by the name of $J$-theory, developed around V.P.~Potapov starting in the 1950s~\cite{Potapov1955,Efimov1973,PotapovMemorialVolume:1994}.
Some of the essential mathematical ideas have been reviewed in Refs.~\cite{Bultheel:1998,Katsnelson:2007}.
In order to keep the present work self-contained, it is useful to state briefly the general spirit and conceptual basis of the methods used by Kovalishina's $J$-theory solution to interpolation problems~\cite{Kovalishina1984}.

Katnelson has described the central idea of the $J$-theory approach to be transforming an interpolation problem into an inequality about analytic functions in some chosen domain~\cite{Katsnelson:2007}.
The approach leads to a Fundamental Matrix Inequality (FMI) for each problem.
Solution of the original interpolation problem then proceeds in two steps: establishing the equivalence with the FMI and then solving the FMI.
\cref{sec:technical} reviews construction of the FMI for the Hamburger moment problem.

\subsection{Hamburger moment problem: general solution}

After formulating the Hamburger moment problem in terms of its FMI, Kovalishina used $J$-theory to construct the general solution for the Stieltjes transform $G(z)$. Kovalishina's result applies only to the nondegenerate case $H \succ 0$, but was later generalized to include the degenerate case \cite{CHEN1998199}.
Kovalishina's result is the following theorem~\cite{Kovalishina1984}:

\begin{theorem}[General solution]\label{thm:general_solution}
The general solution to the non-degenerate $(H \succ 0)$ Hamburger moment problem takes the form
\begin{equation}
\begin{split}
    G(z) = [&\alpha(z) p(z) + \beta(z) q(z)]\\
    &\times [\gamma(z) p(z) + \delta(z) q(z)]^{-1}, 
\end{split}
    \label{eq:hamburger_general_solution}
\end{equation}
where the matrices $\alpha(z), \beta(z),\gamma(z), \delta(z)$ are the component blocks of the matrix
\begin{align}
    \mathfrak{A}(z) =\begin{bmatrix}
        \alpha(z) & \beta(z) \\ \gamma(z) & \delta(z)
    \end{bmatrix}.
\end{align}
and where the arbitrary matrices $p(z)$ and $q(z)$ form a nonsingular positive holomorphic pair.
\end{theorem}
Matrix-valued functions $p(z)$ and $q(z)$ are said to form a nonsingular positive holomorphic pair when both of the conditions
\begin{align}
    \left[ p^\dag(z), q^\dag(z) \right] 
    \begin{bmatrix}
        p(z) \\ q(z)
    \end{bmatrix} &\succ 0,\\
    \left[ p^\dag(z), q^\dag(z) \right] J_2
    \begin{bmatrix}
        p(z) \\ q(z)
    \end{bmatrix} &\succeq 0, \label{eq:pq_J2_positive}
\end{align}
hold. \Cref{sec:technical} sketches the derivation of the general solution using $J$-theory and gives an analytic formula for the coefficient matrix $\mathfrak{A}(z)$ which can be used in numerical implementation.

Not coincidentally, \cref{eq:hamburger_general_solution} has a form very similar to the general solution of the Nevanlinna--Pick problem.
Given this similarity, it is perhaps unsurprising that the full space of solutions can be characterized explicitly.
Since the solution is matrix valued, the characterization is given in terms of Weyl matrix balls.

\subsection{Weyl matrix balls}

Weyl matrix balls are a natural generalization of a disk in the complex plane.
Recall that a disk centered at $c\in\C$ with radius $r>0$ can be parameterized as
\begin{align}
    \disk(c, r) \equiv \{c + rz \mid 1-|z|\geq0\}.
\end{align}
Let $M_{n\times n}(\C)$ denote the space of $n\times n$ complex matrices and suppose $A,B,C\in M_{n\times n}(\C)$.
The Weyl matrix ball with center $C$, left and right radii $A,B$ is defined via
\begin{align}
    \ball(C;A,B) \equiv \{ C + AXB \mid I - X^\dag X \succeq 0 \},
\end{align}
where the condition $I -X^\dagger X \succeq 0$ says that $X$ is a contractive matrix.
In other words, $X$ shrinks the norm of vectors $\vecr$:
\begin{align}
    \norm{X \vecr}^2 = \vecr X^\dag X \vecr \leq \vecr^\dag \vecr = \norm{\vecr}^2.
\end{align}
Equivalently, a matrix $X$ is contractive when its singular values are all less than unity or when $\norm{X}\leq 1$, with $\norm{X}$ denoting the operator norm.
Note that for the scalar case, the Weyl matrix ball reduces to a disk of radius $|AB|>0$ centered at $C$:
\begin{align}
    \ball(C;A,B) = \disk(C,|AB|) && (\text{scalar case }n=1).
\end{align}

\subsection{Bounding properties of Weyl matrix balls}

The general solution to the Hamburger moment problem given in \cref{thm:general_solution} can be characterized geometrically with the help of a positivity condition with respect to the so-called Weyl matrix $W(z)$, which arises naturally when solving the FMI (see \cref{sec:technical}).
The Weyl matrix $W(z)$ may be expressed in terms of the coefficient matrix $\mathfrak{A}(z)$ as
\begin{equation}
W(z) = \mathfrak{A}^{\dag-1}(z) J_2 \mathfrak{A}^{-1}(z)
= \begin{bmatrix}
-R(z) & S(z) \\ S^\dag(z) & -T(z)
\end{bmatrix},
\end{equation}
where the names $R(z)$, $S(z)$, and $T(z)$ for the block components of $W(z)$ follow Ref.~\cite{Kovalishina1984}.
To derive the positivity condition, consider a non-singular positive holomorphic pair $p, q$ and write
\begin{align}
    \begin{bmatrix}
        u \\ v
    \end{bmatrix}
     = \mathfrak{A} \begin{bmatrix}
         p \\q
     \end{bmatrix}.
\end{align}
From \cref{eq:hamburger_general_solution}, the general solution is $G(z) = u v^{-1}$.
By the positivity assumption for $p$ and $q$ (cf. \cref{eq:pq_J2_positive} above),
\begin{align}
    (u^\dag, v^\dag) W  \begin{bmatrix}
        u \\ v
    \end{bmatrix}
    =
    (p^\dag, q^\dag) J_2 \begin{bmatrix}
        p \\ q
    \end{bmatrix}
    \succeq 0 
\end{align}
and thus 
\begin{align}
    (G^\dag, I) W \begin{bmatrix}
        G \\ I
    \end{bmatrix} \succeq 0. 
    \label{eq:Stieltjes_positivity_condition}
\end{align}
\cref{eq:Stieltjes_positivity_condition} is the positivity condition mentioned above.
Kovalishina's theorem relating the general solution in \cref{thm:general_solution} to a Weyl matrix ball can now be stated:
\begin{theorem}[Weyl matrix ball]\label{thm:weyl_matrix_ball}
    The set of matrices $G$ satisfying \cref{eq:Stieltjes_positivity_condition} fills the Weyl matrix ball $\ball(C;\rho_g^{1/2},\rho_d^{1/2})$ with center $C$, left (``gauche") radius $\rho_g$, and right (``droite") radius $\rho_d$:
    \begin{align}
    C &= R^{-1} S \\
    \rho_g &= R^{-1}\\
    \rho_d &= S^\dag R^{-1} S - T.
    \end{align}
\end{theorem}
Kovalishina's proof is sketched in \cref{sec:technical}.
In physical applications, it will be desirable to restrict the bound provided by \cref{thm:weyl_matrix_ball} for the matrix-valued $G(z)$ to bounds for individual components.
The following elementary lemma accomplishes this goal.\footnote{This lemma appears to be a minor novelty of the present work.
However, the basic idea of projection seems sufficiently obvious that the result is likely already known.
}
\begin{lemma}[Projection to a disk]
\label{lemma:weyl-ball-project}
Let $\ball(C; A, B)$ be a Weyl matrix ball with $A,B,C\in M_{n\times n}(\C)$. Let $\vecl, \vecr \in \C^n$ be vectors, and define
\begin{equation}
    \vecl^\dag \ball(C; A, B) \vecr \equiv \{\vecl^\dag M \vecr \mid M \in \ball(C; A, B)\}
\end{equation}
then 
\begin{equation}
\label{eq:weyl-ball-proj}
\vecl^\dag \ball(C; A, B) \vecr = \disk(\vecl^\dag C \vecr, r)
\end{equation}
where $r = ||A^\dag \vecl|| \norm{B\vecr}$.
In other words, \cref{eq:weyl-ball-proj} projects to a disk in $\C$.
\end{lemma}
A proof is given in \cref{sec:technical}.
Letting $\bm{e}_a$ denote a real unit vector in the $a$th direction, one can define a \emph{component-wise Wertevorrat}, 
\begin{align}
    \Delta_{ab}(z) \equiv \bm{e}_a \ball\left(C, \rho_g^{1/2}, \rho_d^{1/2}\right) \bm{e}_b.
    \label{eq:wertevorrat_components}
\end{align}
Together, \cref{lemma:weyl-ball-project} and \cref{thm:weyl_matrix_ball} furnish component-wise bounds for the Stieltjes transform
\begin{align}
    G_{ab}(z) \in \Delta_{ab}(z). \label{eq:stieltjes_bound}
\end{align}
The applicability of \cref{eq:wertevorrat_components,eq:stieltjes_bound} to physical problems is one of the main results of the present work.
The preceding inclusion evaluated at  $z=\lambda+i\epsilon$ provides a bound on the smeared spectral function 
\begin{align}
\begin{split}
    \min \Im \, \partial\Delta_{ab}(\lambda+i\epsilon)
    &< \pi \tilde{\rho}_{ab, \epsilon}(\lambda)\\
    &< \max \Im \, \partial\Delta_{ab}(\lambda + i \epsilon),
\end{split}\label{eq:density_bound}
\end{align}
where $\partial \Delta_{ab}(z)$ denotes the boundary of the Wertevorrat.
Multiplying by $\lambda$ to account for the Jacobian factor in \cref{eq:physical_spectral_function} gives a similar bound for the physical spectral function.
In applications, these bounds can be evaluated numerically.

\section{Relationship to other methods\label{sec:other_methods}}

Approaches to spectral reconstruction problems in lattice field theory are manifold.
Correlated fitting with Bayesian constraints 
is a popular and versatile approach in many contexts~\cite{Michael:1993yj,Lepage:2001ym,Morningstar:2001je}.
Solution of a generalized eigenvalue problem (GEVP) using a variational set of operators provides access to excited states (see Refs.~\cite{Luscher:1990ck,Blossier:2009kd} and references therein).
Variations on Prony's method have also been proposed~\cite{Cushman:2019tcv,Fischer:2020bgv,Fleming:2023zml,Fleming:2009wb}.
The Lanczos algorithm has been the subject of much recent activity
both in lattice field theory
~\cite{Wagman:2024rid,Hackett:2024nbe,Hackett:2024xnx,Chakraborty:2024exj,Ostmeyer:2024qgu,Abbott:2025yhm} and beyond~\cite{Parker:2018yvk,Nandy:2024evd,Dodelson:2025rng,Beetar:2025erl}.
The main theoretical bounds for energy eigenvalues come from the variational principle, the exponential convergence properties of the GEVP~\cite{Luscher:1990ck,Blossier:2009kd}, and Cauchy's interlacing theorem~\cite{Hwang:2024,Horn:1985,doi:10.1137/1.9781421407944,Hwang:2024}, as well as more recently Kaniel--Paige--Saad (KPS) convergence theory~\cite{Kaniel:1966,Paige:1971,Saad:1980,Wagman:2024rid,Hackett:2024nbe,Abbott:2025yhm}.

Interest in smeared spectral reconstructions increased dramatically following work by Hansen, Lupo, and Tantalo~\cite{Hansen:2019idp}, which has connections to Gaussian processes~\cite{Horak:2021syv,DelDebbio:2024lwm}.
Methods from Nevanlinna--Pick interpolation~\cite{Nicolau2016,Nevalinna1919,Nevanlinna1929,Pick1915,PickInterpolationBook} provide a complementary picture and, when the problem is formulated in frequency space~\cite{PhysRevLett.126.056402,Fei_2021}, furnish rigorous uncertainty bounds for smeared spectral functions~\cite{Bergamaschi:2023xzx}.

The remainder of this section focuses on the connections to the Rayleigh--Ritz method and the Lanczos algorithm.

\subsection{The Rayleigh--Ritz method\label{sec:RayleighRitz}}

The Rayleigh--Ritz (RR) method~\cite{Ritz+1909+1+61,LEISSA2005961} for approximating eigenvalues has been reviewed recently
by Ref.~\cite{Abbott:2025yhm} in the context of spectroscopy for lattice field theory.
The Rayleigh--Ritz method approximates eigenvalues of the transfer matrix by restricting the problem to a finite-dimensional subspace and performing a min-max procedure on the Rayleigh quotient over that subspace, where the problem can then be formulated as generalized eigenvalue problem.
The resulting eigenvalues are known as Ritz values.

The Rayleigh--Ritz method is closely related to the moment problems discussed above.
Suppose that the moments $C_0, \dots, C_{2n-1}$ are given.
Define a rational function of degree $(n-1,n)$ by
\begin{equation}
    r(z)
    = \frac{p_0 + p_1 z + \dots + p_{n-1} z^{n-1}}
    {q_0 + q_1 z + \dots + q_{n-1} z^{n-1} + z^n},
    \label{eq:RR-rational-function}
\end{equation}
where the coefficients can be chosen, for instance via Pad{\'e} approximation~\cite{Press:2007ipz}, such that
\begin{equation}
\label{eq:rational-asymptotic}
    -r(z) = \frac{C_0}{z} + \dots + \frac{C_{2n-1}}{z^{2n}}
    + O(1/z^{2n+1}).
\end{equation}
A partial fraction decomposition applied to \cref{eq:RR-rational-function} can be used to write
\begin{equation}
    -r(z) = \sum_{k=1}^n \frac{A_k}{z - \lambda_k}
    =
    \sum_{m=0}^\infty
    \frac{\sum_{k=1}^n A_k \lambda_k^m}{z^{m+1}} \label{eq:partial_fraction}
\end{equation}
for some $\lambda_k$ and $A_k$ and where the second equality follows from Taylor expansion around $z=\infty$.
Comparison with \cref{eq:rational-asymptotic} shows that $A_k$ and $\lambda_k$ are related to the moments via
\begin{equation}
    \label{eq:sum-of-exp}
    C_t = \sum_{k=1}^n A_k \lambda_k^t.
\end{equation}
On the other hand, Ref.~\cite{Abbott:2025yhm} has shown that, given \cref{eq:sum-of-exp}, the expansion parameters $\lambda_k$ appearing in the partial-fraction decomposition \cref{eq:partial_fraction} can be identified with the Ritz values and the $A_k$ are related to the spectral weights.

The connection between the Rayleigh--Ritz method and moment problem may therefore be summarized as follows. 
The Rayleigh--Ritz method gives the \emph{unique} rational function of degree $(n-1,n)$ that interpolates the Euclidean-time correlator $C_t$ with an $n$-state approximation to the transfer matrix.
On the other hand, the general solution of the moment problem gives a full characterization of all possible solutions, without imposing saturation by $n$ states.
In applications, the utility of a particular solution compared to the general solution must be judged on a case-by-case basis.

\subsection{The Lanczos algorithm}

The Lanczos algorithm is an iterative method for computing eigenvalues.
Ref.~\cite{Abbott:2025yhm} discussed the equivalence of the Lanczos algorithm~\cite{Lanczos:1950zz} to the Rayleigh--Ritz method.
Building on \cref{sec:RayleighRitz}, a direct connection also exists to moment problems. 
The remainder of this section reviews this connection as described in Chapters 5 and 6 of Ref.~\cite{schmudgen2017moment} before offering a new physical interpretation.

Suppose as before that moments $C_0, C_1, \dots$ are given.
Then the vector space $\mathcal{H}_C = \C[x]$ can be given a Hilbert space structure by defining the inner product
\begin{align}
    \braket{f | g }
    = \int dx\, \rho(x) f^*(x) g(x)
    = L_C\left[f^*(x) g(x) \right]
\end{align}
where $L_C : \C[x] \to \C$ is the linear functional defined by $L_C[x^n] = C_n$.
Application of the Gram--Schmidt procedure with respect to to the set $\{1, x, x^2, \dots\}$ under this inner product allows one to construct a sequence of orthogonal polynomials $p_n$ with $p_0(x)=1$ and $\braket{p_n | p_m} = \delta_{nm}$.
The orthogonal polynomials satisfy a three-term recurrence relation, which takes the form
\begin{equation}
    x p_n(x) = a_n p_{n+1}(x) + b_n p_{n}(x) + a_{n-1} p_{n-1}(x),\label{eq:lanczos_recurrence}
\end{equation}
where $a_n$ and $b_n$ are coefficients that can be computed from the moments.
\Cref{eq:lanczos_recurrence} is precisely the recurrence relation that defines the Lanczos algorithm applied to the operator $X : \mathcal{H}_C \to \mathcal{H}_C$ given by $X(f(x)) = x f(x)$ with initial vector $p_0(x)=1$. 
The Lanczos procedure on $\C[x]$ can be related to the Lanczos procedure on the physical Hilbert space via the embedding $x^n \mapsto \hat{T}^n \ket{\psi}$.
This embedding provides a connection to the operator-theoretic view of the moment problem which makes a precise connection between solutions to the moment problem and self-adjoint extensions of the operator $X$.

From a physical perspective, this mathematically well-known correspondence states that solutions to a given moment problem arise exactly from embeddings of $\mathcal{H}_C$ into a physical Hilbert space such that the transfer matrix $\hat{T}$ extends the operator $X$.
The physical interpretation of the present work makes precise the statement in Ref.~\cite{Abbott:2025yhm} that any problem involving reconstructing a weighted sum of exponentials like \cref{eq:sum-of-exp} with $\lambda_k = e^{-E_k}$ can be thought of as arising from \emph{some} Hilbert space.

%%%%%%%%%%%%%%%%%%%%%%%%%%%%%%%%%%%%%%%%%%%%%%%%
\section{Numerical results \label{sec:numerical_results}}
%%%%%%%%%%%%%%%%%%%%%%%%%%%%%%%%%%%%%%%%%%%%%%%%

\subsection{Generic spectral reconstruction}

\begin{figure}
    \centering
    \includegraphics[width=1\linewidth]{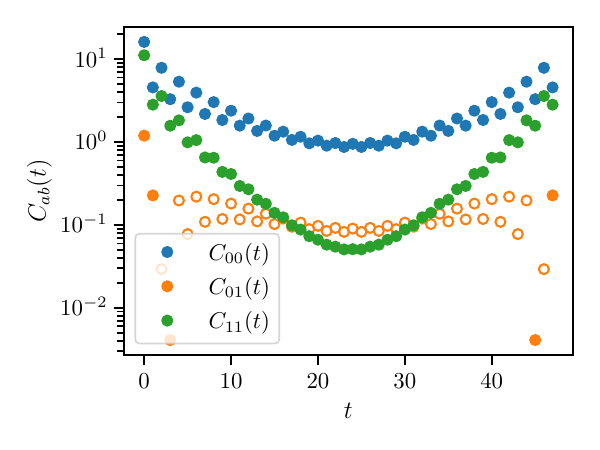}
    \caption{The distinct components of the $2\times2$ matrix-valued correlation function in \cref{eq:staggered_example}.
    Open symbols denote points in which the off-diagonal component of the correlator becomes negative.
    }
    \label{fig:staggered_correlator_2x2}
\end{figure}

Consider a generic $2\times 2$ matrix-valued staggered-fermion correlation function of the form
\begin{align}
\begin{split}
    &C_{ab}(t) = \sum_{n=0}^{N_{\rm max}} \frac{Z_{na}^* Z_{nb}}{2 E_n} (e^{-E_n t} + e^{-E_n(\beta-t)})\\
    &\quad+\sum_{m=0}^{M_{\rm max}} \frac{Z_{ma}^{{\rm osc}*} Z_{mb}^{\rm osc}}{2E_m^{\rm osc}} (-1)^t (e^{-E_m^{\rm osc} t} + e^{-E_m^{\rm osc}(\beta-t)}).
\end{split}\label{eq:staggered_example}
\end{align}
For simplicity, consider the case in which the spectral weights coincide in the decaying and oscillating towers of states: $Z_{na} = Z_{na}^{\rm osc}$. 
For concreteness, take a temporal extent of $\beta=48$ with
$N_{\rm max}=M_{\rm max}=20$ states and with numerical values for the energies and spectral weights given by
\begin{align}
    E_n = \tfrac{1}{10}(n+1),  && 
    E_m^{\rm osc} = \tfrac{1}{5}(m+1)
\end{align}
\begin{align}
Z_{n0} &=
\begin{cases}
1 & n \text{ even} \\
\tfrac{1}{10} & n \text{ odd},
\end{cases}  && 
Z_{n1} =
\begin{cases}
-\tfrac{1}{10} & n = 0 \\
+1 & n \text{ odd} \\
+\tfrac{1}{10} & n>0 \text{ even}.
\end{cases}
\end{align}
Apart from the presence of the oscillating states, essentially the same system was used as an example in Ref.~\cite{Hackett:2024nbe}.
\Cref{fig:staggered_correlator_2x2} shows the distinct components of the matrix-valued correlator.

\begin{figure*}
    \centering
    \includegraphics[width=0.48\linewidth]{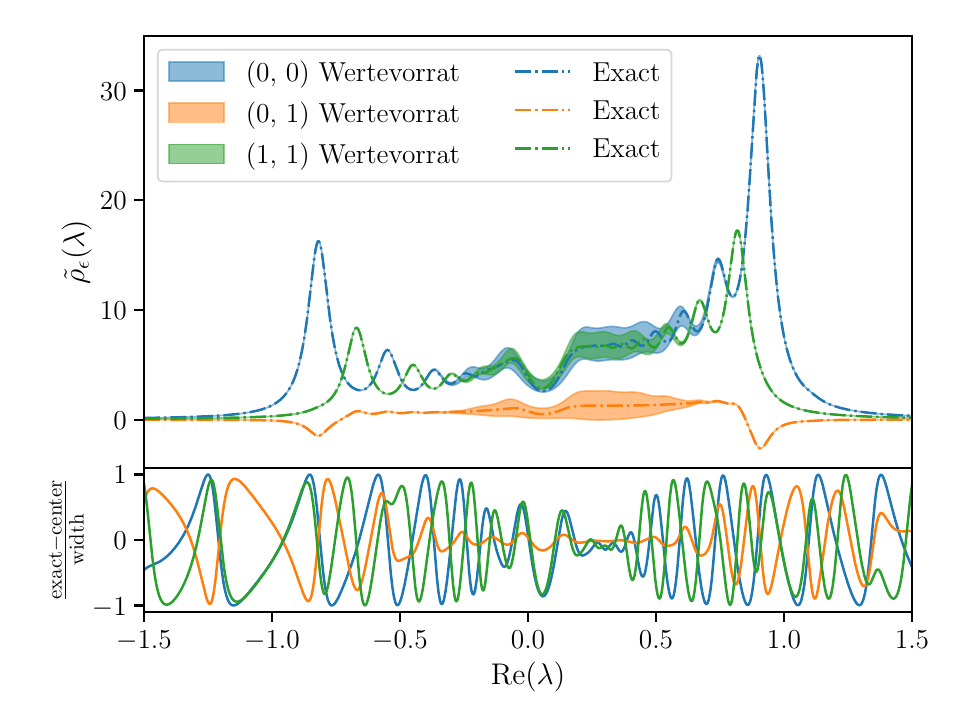}
    \includegraphics[width=0.51\linewidth]{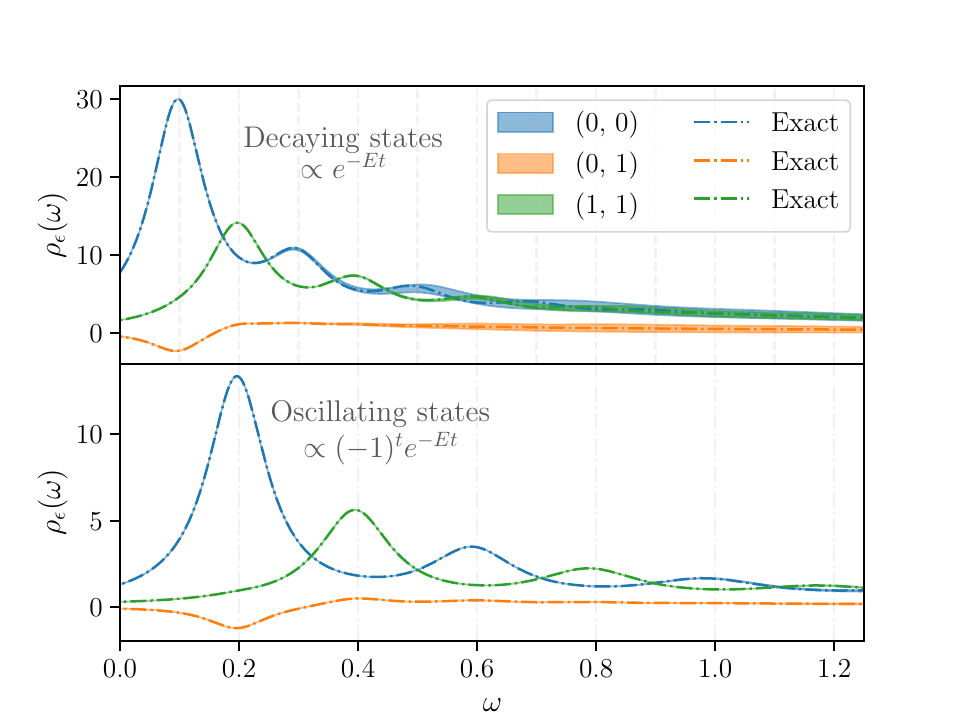}
    \caption{\textbf{Left:} Smeared spectral reconstruction $\tilde{\rho}_\epsilon(\lambda) = \tfrac{1}{\pi} \Im G(\lambda+i\epsilon)$ in the complex-$\lambda$ plane
    for the $2\times2$ matrix-valued correlator defined in \cref{eq:staggered_example} evaluated at $\epsilon = 0.05$
    using all $\beta=48$ moments.
    In the upper panel, 
    the colored bands denote uncertainty coming from the Wertevorrat for each component, as defined in \cref{eq:wertevorrat_components}.
    The dotted lines show the exact answer.
    The bottom panel shows the normalized location of the exact answer within the bounding set of Wertevorrat.
    As required, the normalized location lies within the interval $(-1,1)$.
    \textbf{Right:}
    The smeared spectral reconstruction $\rho_\epsilon(\omega)$ in the complex-energy plane,
    constructed from the results on the left using 
    $\rho_\epsilon(\omega) = \lambda(\omega) \tilde{\rho}_\epsilon(\lambda(\omega))$.
    The upper and lower panels shows the spectral functions for the decaying and oscillating states, respectively.
    In all cases, the exact answer lies with in the shaded bound from the Wertevorrat.
    }
    \label{fig:staggered_reconstruction}
\end{figure*}

The left-hand side of \Cref{fig:staggered_reconstruction} shows a smeared spectral reconstruction of \cref{eq:staggered_example} in the complex-$\lambda$ plane at fixed $\epsilon=0.05$ above the real-$\lambda$ line.
Since \cref{eq:staggered_example} corresponds to the Hamburger moment problem, the density $\tilde{\rho}_{ab}(\lambda)$ has support on the real-$\lambda$ line.
The shaded bands are the bounds of the Wertevorrat from \cref{eq:density_bound}, and the dotted lines are the exact answer.
The lower left of \cref{fig:staggered_reconstruction} shows the normalized location of the exact answer within the Wertevorrat, 
\begin{align}
    \rm{normalized~location} = \frac{\rm{exact-center}}{\rm width}.
\end{align}
The fact that this quantity always lies between $-1$ and $1$ (as required by \cref{eq:density_bound}) emphasizes that the Wertevorrat is a convex bounding set and \emph{not} a statistical confidence interval.

The effects of thermal states with $|\lambda|>1$ are severely suppressed in \cref{fig:staggered_reconstruction}, as expected from $\rho_{ab}(\lambda) \propto \lambda^{-\beta}$ in this region.
Overall, these results provide a quantitative example of the general qualitative picture shown in \cref{fig:eigenvalue_plane}.

The right side of \cref{fig:staggered_reconstruction} show the smeared spectral functions  $\rho_\epsilon(\omega)=\lambda(\omega) \tilde{\rho}_\epsilon(\lambda(\omega))=\lambda(\omega) \tfrac{1}{\pi}\imag G(\lambda(\omega)+i\epsilon)$ associated with the decaying and oscillating states in \cref{eq:staggered_example}, respectively.
These results are obtained using \cref{eq:physical_spectral_function}.
In practical lattice QCD calculations, it may be important to monitor the quantitative size of ``bleed-in effects" from adjacent eigenvalues domains on the real-$\lambda$ line, i.e., the influence of thermal states ($\lambda > 1$) at low energy and from staggered oscillating states ($\lambda < 0$) at high energy.
It is perhaps useful to note that the different $\lambda$ regions can readily be exchanged,
\begin{align}
C(t) &\mapsto C(t-\beta) && |\lambda|<1 \longleftrightarrow |\lambda|>1   \\
C(t) &\mapsto (-1)^t C(t) && \lambda >0 \longleftrightarrow \lambda < 0,
\end{align}
which provide a set of consistency conditions.

\subsection{Scaling properties}

Consider a matrix-valued  correlation function of the form 
\begin{align}
     C_{ab}(t) =& \sum_{n=0}^{N_{\rm max}} Z_{na}^* Z_{nb} e^{-E_n t}, \label{eq:NxN_matrix_correlator}
\end{align}
which for simplicity is taken to contain decaying contributions only.
Suppose the energies and spectral weights are given by
\begin{align}
    E_n &= E_0 (n+1)\\
    Z_{na} &= J_a(5 E_n),
\end{align}
where $J_a(x)$ are the Bessel functions of the first kind.
This model is chosen to provide a general $N\times N$ matrix-valued correlator with spectral weights that peak in different regions.
For concreteness, take $E_0 =\tfrac{1}{100}$ and suppose that $N_{\rm max}=200$.

\begin{figure*}
    \centering
    \includegraphics[width=0.75\linewidth]{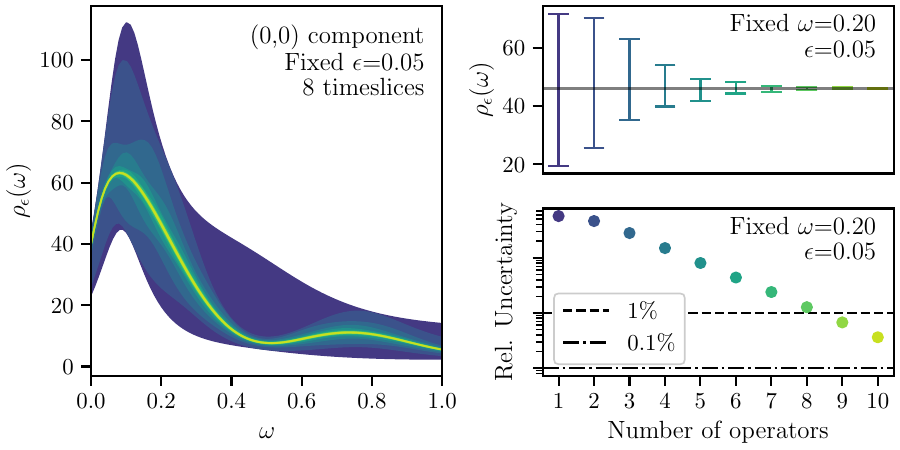}
    \includegraphics[width=0.06\linewidth]{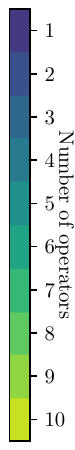}
    \caption{
    Scaling of bounds for the (0,0) component of the smeared spectral function 
    $\rho_\epsilon(\omega) = \lambda(\omega) \tfrac{1}{\pi}\imag G(\lambda(\omega)+i\epsilon)$
    with the size of the input $N\times N$ matrix-valued correlation function defined in \cref{eq:NxN_matrix_correlator}.
    The left panel shows the smeared spectral function reconstructed from 8 times at fixed $\epsilon=0.05$ above the real-$\lambda$ line for $N\in\{1,2,\dots 10\}$ operators. 
    On the right, the upper and lower panels show the convergence and relative uncertainty, respectively, as the number of operators increases.
    In the upper panel, the horizontal line shows the location of the exact result.
    The relative uncertainty decreases exponentially with the number of operators.
    }
    \label{fig:varying_operators}
\end{figure*}

\Cref{fig:varying_operators} shows the scaling of the fractional uncertainty in the reconstruction of the (0,0) component of the spectral density $\rho_\epsilon(\omega) = \lambda(\omega) \tfrac{1}{\pi} \imag G(\lambda(\omega) + i \epsilon)$ as the size of the matrix $N\times N$ correlator matrix is increased with $N\in\{1,2,\dots 10\}$ using 8 Euclidean times, $\omega= 0.2$, and $\epsilon=0.05$.
The uncertainty is observed to decrease roughly exponentially with $N$.

\begin{figure}
    \centering
    \includegraphics[width=1.0\linewidth]{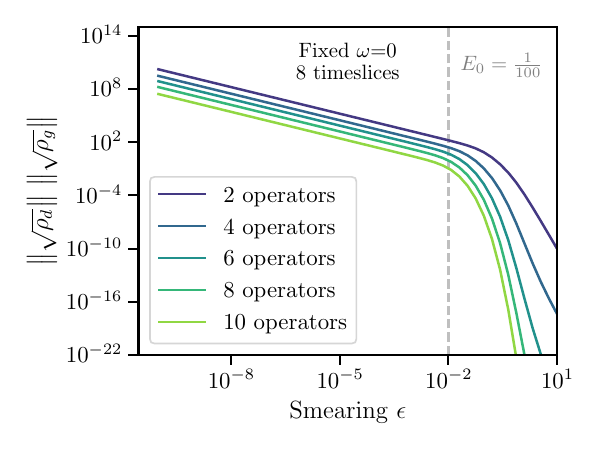}
    \caption{
    Uncertainty in the Stieltjes transform associated with \cref{eq:NxN_matrix_correlator}, measured by the diameter of the Weyl matrix ball as the smearing width varies.
    Uncertainty increases as the smearing decreases.
    The vertical line denotes the location of the parameter $E_0=\tfrac{1}{100}$ in \cref{eq:NxN_matrix_correlator}.
    }
    \label{fig:epsilon_scaling_staggered_model}
\end{figure}

\Cref{fig:epsilon_scaling_staggered_model} shows uncertainty on the Stieltjes transform at a distance $\epsilon$ above the real-$\lambda$ line and for $\lambda=1$ (i.e., $\omega=0$).
The uncertainty is measured by the diameter of the Weyl matrix ball, which is quantified by the product $\norm{\sqrt{\rho_d}} \norm{\sqrt{\rho_g}}$.
As expected, the uncertainty increases as the smearing decreases.
At least for the particular example, two power-law scaling regions are present, one for $\epsilon \gg E_0$ and one for $\epsilon \ll E_0$.
A more detailed understanding of generic scaling properties as a function of $\epsilon$ and the large-energy behavior of the spectral function may be useful in practical applications.
Such investigation exceeds the scope of the present work.

%%%%%%%%%%%%%%%%%%%%%%%%%%%%%%%%%%%%%%%%%%%%
\section{Considerations regarding finite precision\label{sec:noise}}
%%%%%%%%%%%%%%%%%%%%%%%%%%%%%%%%%%%%%%%%%%%%

The numerical examples considered in \cref{sec:numerical_results}
were computed using exact numerical inputs.
As seems to be necessary with other methods for smeared spectral reconstruction (e.g., Ref.~\cite{Hansen:2019idp,Bergamaschi:2023xzx}), intermediate steps were carried out using extended-precision arithmetic~\cite{mpmath}.\footnote{Numerical examples in this work used mpmath~\cite{mpmath} with 200 digits of precision.}
One can ask about how the problem changes for input data that are noisy or, equivalently, known to finite numerical precision.
As discussed in \cref{sec:solution_existence}, a necessary condition for the existence of a solution to a moment problem is a non-negative Hankel matrix $\hankel\succeq 0$.
For a fixed numerical precision, this condition can be expected eventually to fail as the size of the input data (the number of moments and/or the size of the matrix blocks) increases.

Imposing the condition $\hankel\succeq 0$ on noisy Monte-Carlo data through a projection step has been advocated in Ref.~\cite{Yu_2024}.
A scheme in this spirit is attractive for several reasons.
First, the positivity condition is a non-perturbative statement of causality, and therefore imposing it therefore does not introduce any model assumptions.
Second, it offers advantages compared to the Pick criterion appearing Nevanlinna--Pick interpolation, namely that the Hankel matrix is defined in terms of the Euclidean-time correlation function $C_{ab}(t)$ and that the entries of the Hankel matrix are linear in the correlation function.
Third, the algorithm of Ref.~\cite{Yu_2024} elegantly imposes $\hankel\succeq 0$ by iteratively projecting onto the space of positive semi-definite matrices and then projecting onto the space of Hankel matrices.
As Ref.~\cite{Yu_2024} emphasizes, this projection is made possible by the fact that both the space of Hankel matrices and the space of positive semi-definite matrices are convex.
Numerical methods in convex optimization are well developed~\cite{Press:2007ipz}.

An alternative perspective on noisy data has been given in Refs~\cite{Wagman:2024rid,Hackett:2024nbe} in the context of the Lanczos algorithm.
When the Lanczos algorithm is implemented using finite-precision arithmetic, numerical artifacts know as ``spurious eigenvalues" can arise.
A well-known approach for identifying and removing spurious eigenvalues is the Cullum--Willoughby (CW) method~\cite{Cullum:1981,Cullum:1985}.
The CW method works by looking for stability of eigenvalues under removal of initial vectors (i.e., the vectors excited by the operators) from the Krylov space; true eigenvalues are sensitive to this removal, while the spurious eigenvalues are not.
Ref.~\cite{Hackett:2024nbe} introduced a new and closely related method called the ZCW test, which identifies spurious eigenvalues as those with unphysically small overlaps onto initial states.
Either method can be used as part of a projection step to remove acausal fluctuations.

Theoretical rigor requires that calculations in lattice QCD be systematically improvable.
In particular, this requirement means that any projection step to impose $\hankel\succeq 0$ must not introduce an unquantified bias to the spectral reconstruction.
Since correlations are physical in the sense that they persist in the limit of infinite Monte-Carlo statistics, a necessary condition seems to be that the projection step incorporate and respect correlations to the greatest extent consistent with causality.
Further study seems to be necessary to determine the degree to which such correlations can be preserved by projection steps like the algorithm in Ref.~\cite{Yu_2024} or the CW and ZCW methods.

Questions of numerical precision have also been considered in related problems of analytic continuation.
Ref.~\cite{Costin:2022hgc} has studied properties of analytic continuations using Pad{\'e} and conformal-map approximants in the case where expansion coefficients are known only with finite precision or are subject to noise.
One of the main results was that noisy inputs lead to the breakdown of the Pad{\'e} approximant at an order which is proportional to logarithm of the noise strength.
In other words, linearly increasing the order of a Pad{\'e} approximant requires exponential improvement in the precision of the inputs.

\section{Conclusions \label{sec:conclusions}}

Moment problems have a long history of study within the mathematical literature.
This work has shown that the calculation of smeared spectral functions from Euclidean-time correlation functions can be viewed as a Hamburger moment problem.

Solutions to the Hamburger moment problem can be phrased in terms of the Stieltjes transform $G(z)$, \cref{eq:stieltjes-trans-def}, which is interpreted physically as the familiar K{\"a}ll{\'e}n--Lehmann spectral representation.
Kovalishina has computed the general solution of the matrix-valued truncated Hamburger moment problem, giving a bound for Stieltjes transform at each point $z\in \C/\R$.
The present work gives the first numerical realization of Kovalishina's rigorous bounds.

The basic inputs are matrix-valued correlation functions $C_{ab}(t)$.
Such matrices are already computed as a matter of course in lattice field theory, for instance, in calculations of scattering amplitudes based on the finite-volume formalism~\cite{Luscher:1990ux,Briceno:2017max}, which require knowledge of a large number of finite-volume energy levels.
These energy levels are typically computed by solving a generalized eigenvalue problem to extract $N$ levels given an interpolating set with $N$ operators, i.e., an $N\times N$ matrix.
Within the present work, numerical evidence suggests that reliable \emph{smeared} spectral knowledge remains accessible above the $N$th level and that uncertainties decrease exponentially as the number of operators increases.
Since many correlation functions in lattice field theory are plagued by an exponentially growing noise-to-signal ratio, such scaling may also be of practical interest elsewhere.

A complementary approach to scattering amplitudes, put forward by Hansen and Bulava~\cite{Bulava:2019kbi}, is based on using smeared spectral functions to implement the $i\epsilon$ prescription for Green functions in QFT.
The present work implements precisely this smearing by evaluating the Stieltjes transform $G(z)$ at $z=\omega+i\epsilon$.
Within this context, even if one is primarily interested in some particular scalar-valued smeared spectral function, the present results suggest a potentially large benefit to be gained by utilizing a large variational set of operators.
For example, suppose one is interested in amplitudes associated with correlation functions of the form
\begin{align}
    \bra{\psi} J(t) J(0) \ket{\psi},
\end{align}
where $J$ might be an external weak or electromagnetic current and $\ket{\psi}$ is some hadronic state like the vacuum or a heavy meson.
Mirroring what is already done in the finite-volume formalism, one can imagine embedding the current $J$ within a larger set of multihadron interpolating operators that overlap strongly with, say, $2\pi$ or $3\pi$ states in order to realize the scaling favorable scaling seen in \cref{fig:varying_operators}.

Although the primary focus of the present work has been the theoretical problem in which the Euclidean-time correlation function is taken to be known exactly, concerns about finite numerical precision are crucial for practical applications.
This question has received some attention in the recent literature~\cite{Yu_2024,Abbott:2025yhm,Costin:2022hgc} and deserves further investigation.

\section*{Acknowledgements}

We gratefully acknowledge useful conversations with 
Tom Blum,
Will Detmold,
Daniel Hackett,
Luchang Jin,
Ethan Neil,
Miguel Salg,
Douglas Stewart,
Fernando Romero-L{\'o}pez,
Julian Urban, and
Mike Wagman.

The numerical calculations in this work made use of \textsc{NumPy}~\cite{vanderWalt:2011bqk,Harris:2020xlr} and \textsc{mpmath}~\cite{mpmath}.
Figures were generated using \textsc{matplotlib}~\cite{Hunter:2007} and \textsc{seaborn}~\cite{Waskom2021}.

PO is supported in part by the U.S. Department of Energy, Office of Science, Office of Nuclear Physics under grant Contract Number DE-SC0012704 (BNL).
RA is supported
%in part by the National Science Foundation under Cooperative Agreement PHY-2019786 (The NSF AI Institute for Artificial Intelligence and Fundamental Interactions, http://iaifi.org/),
in part by the U.S.~Department of Energy, Office of Science, Office of Nuclear Physics under grant Contract Number DE-SC0011090, as well as 
by the U.S.~Department of Energy SciDAC5 award DE-SC0023116 and the High Energy Physics Computing Traineeship for Lattice Gauge
Theory (DE-SC0024053).

\appendix

\section{Technical results\label{sec:technical}}

This appendix collects various technical elements related to the formulation of the Hamburger moment problem as interpolation problem and its solution using $J$-theory.
The contents are structured as follows:
\begin{itemize}
    \item \cref{sec:lemma_block_matrices} states the Fundamental Lemma for Block Matrices, 
    \item \cref{sec:FMI_formulation} discusses formulation of the the Fundamental Matrix Inequality,
    \item \cref{sec:FMI_solution} discusses the solution of the Fundamental Matrix Inequality using $J$-theory,
    \item \cref{sec:Weyl_ball} sketches the proof of Kovalishina's \cref{thm:weyl_matrix_ball}, and
    \item \cref{sec:projection_lemma} proves \cref{lemma:weyl-ball-project}.
\end{itemize}

\subsection{Fundamental Lemma on Block Matrices\label{sec:lemma_block_matrices}}

Potapov and Efimov have discussed the importance of the following lemma for solving block inequalities~\cite{Efimov1973}:
\begin{lemma}[Fundamental lemma on block matrices]\label{lemma:block_matrices}
Suppose a Hermitian matrix 
\begin{align}
    \begin{bmatrix}
        \mathcal{A} & \bm{\mathcal{B}}\\
        \bm{\mathcal{B}}^\dagger & \mathcal{C}
    \end{bmatrix}
\end{align}
is split into blocks such that $\mathcal{A}$ and $\mathcal{C}$ are square matrices.
The matrix is non-negative if and only if the following conditions are satisfied:
\begin{enumerate}
    \item $\mathcal{A} \succeq 0$
    \item $\mathcal{A} X = \bm{\mathcal{B}}$ has at least one solution
    \item $\mathcal{C}-X^\dag \mathcal{A} X\succeq 0$ holds, where $X^\dag \mathcal{A} X$ does not depend on the choice of $X$.
\end{enumerate}
\end{lemma}
A proof of \cref{lemma:block_matrices} is also provided in Appendix A4 of Ref.~\cite{schmudgen2017moment}.
If $\mathcal{A}$ is further assumed to be nonsingular, then the equation $\mathcal{A} X = \bm{\mathcal{B}}$ has the unique solution $X = \mathcal{A}^{-1} \bm{\mathcal{B}}$, and the final condition in \cref{lemma:block_matrices} takes the form
\begin{equation}
    \mathcal{C} - \bm{\mathcal{B}}^\dagger \mathcal{A}^{-1} \bm{\mathcal{B}} \succeq 0. \label{eq:block_lemma_A_inv}
\end{equation}

\subsection{Fundamental Matrix Inequality\label{sec:FMI_formulation}}

The Fundamental Matrix Inequality (FMI) for a classical interpolation problem has the form~\cite{Kovalishina1984,Katsnelson:2007}
\begin{align}
    \begin{bmatrix}
        \mathcal{A} & \bm{\mathcal{B}}(z) \\
        \bm{\mathcal{B}}^\dag(z) & \mathcal{C}(z)
    \end{bmatrix} \succeq 0. \label{eq:FMI}
\end{align}
The ``information block" $A$ in the upper-left is a Hermitian matrix which involves the interpolation data only; the interpolation problem has a solution if and only if $A\succeq 0$.
The ``interpolation block-vector" $\bm{\mathcal{B}}(z)$ specifies the interpolation conditions, involving both unknown function and the interpolation data itself.
Finally, the ``class block" $\mathcal{C}(z)$ contains the unknown function only and controls the function space to which it belongs.

The FMI acquires a particular form in the concrete case of the truncated Hamburger moment problem with inputs $C_0, C_1, \dots C_{2n}$, from which one seeks to determine the Stieltjes transform $G(z)$.
The information, interpolation, and function-class blocks appearing in the FMI are given by
\begin{align}
    \mathcal{A} = \hankel
    &= \begin{bmatrix}
        C_0 & C_1 & \dots & C_n\\
        C_1 & C_2 & \dots & C_{n+1}\\
        \vdots & \vdots & \ddots & \vdots\\
        C_n & C_{n+1} & \dots & C_{2n}
    \end{bmatrix}
    , \\
    \bm{\mathcal{B}}(z) &= \bm{b}(z) G(z) - \bm{c}(z)\\
    \mathcal{C}(z) &= \frac{G(z)-G^\dag(z)}{z-z^*}.
\end{align}
As indicated, $H$ is understood to be a block $n\times n$ matrix constructed from the moments from $C_0, C_1, \dots C_{2n}$.
Dimensional consistency requires that the block vectors $\bm{\mathcal{B}}$, $\bm{b}$, and $\bm{c}$ each have $n$ components.
The lower-right component $\mathcal{C}(z)$ is a $1\times 1$ block.
The block vectors $\bm{b}(z)$ and $\bm{c}(z)$ appearing in the interpolation block are given by
\begin{align}
    \bm{b}(z) &= \begin{bmatrix}
        I \\
        z I \\ 
        z^2 I \\
        \vdots \\
        z^{n-1}I \end{bmatrix},
\end{align}
\begin{align}
    -\bm{c}(z) &= \begin{bmatrix}
        0\\
        C_0\\
        C_1 + z C_0\\
        \vdots\\
        C_{n-2} + z C_{n-3}+\dots + z^{n-2} C_0
    \end{bmatrix}.
\end{align}
One sees that $\bm{\mathcal{B}}(z)$ would be singular at infinity unless the interpolation condition \cref{eq:hamburger_interpolation_condition} is satisfied.
When $\mathcal{A}$ is non-singular, \cref{eq:FMI} reduces to \cref{eq:block_lemma_A_inv}.

For the Hamburger moment problem, Kovalishina has shown that FMI in the form of \cref{eq:block_lemma_A_inv} may be written as~\cite{Kovalishina1984}
\begin{align}
    \left[ G^\dagger(z), I\right] \frac{W(z^*,z)}{(z-z^*)/i}
    \begin{bmatrix}
        G(z) \\ I
    \end{bmatrix} \succeq 0.
    \label{eq:FMI_Weyl}
\end{align}
The matrix $W(z^*,z)$ in the numerator is known as the Weyl matrix and takes the form
\begin{align}
    W(z_1,z_2) = J_2 + J_2 \mathscr{H}(z_1,z_2) J_2,
\end{align}
where 
$J_2 = \left[\begin{smallmatrix} 0 & i I\\ -i I & 0 \end{smallmatrix}\right]$ and
\begin{align}
\begin{split}
\mathscr{H}(z_1, &z_2) = 
\frac{z_1-z_2}{i}\\
&\times\begin{bmatrix}
    -\bm{c}^\dagger(z^*_1) \\
    \bm{b}^\dag(z^*_1)
\end{bmatrix}
\hankel^{-1} 
\left( -\bm{c}(z_2), \bm{b}(z_2) \right)
\end{split}
\end{align}
The Weyl matrix marks the first appearance the eponymous matrices of $J$-theory.
The structure of other interpolation problems is generically reflected in the appearance of different $J$-matrices ($J\neq J_2$) satisfying $J = J^\dag$ and $J^2 = 1$.

\subsection{Solution of the fundamental matrix inequality\label{sec:FMI_solution}}

This section sketches the ideas leading to Kovalishina's solution of the FMI~\cite{Kovalishina1984}.
One begins with the \emph{Ansatz}
\begin{align}
    \begin{bmatrix}G(z) \\ I \end{bmatrix}
    &= \mathfrak{A}(z)
    \begin{bmatrix} p(z) \\ q(z) \end{bmatrix}, \label{eq:ansatz}\\
    \mathfrak{A}(z)
    &= \begin{bmatrix}
        \alpha(z) & \beta(z) \\ \gamma(z) & \delta(z)
    \end{bmatrix},
\end{align}
where $\mathfrak{A}(z)$ is a coefficient matrix and $p(z), q(z)$ form a nonsingular positive holomorphic pair (see \cref{eq:pq_J2_positive}).
With this Ansatz, the solution $G(z)$ can then be written as a M{\"o}bius transformation:
\begin{align}
\begin{split}
G(z) = [&\alpha(z) p(z) + \beta(z) q(z)]\\
    &\times [\gamma(z) p(z) + \delta(z) q(z)]^{-1}.
\end{split}
\end{align}

To construct a solution, one must relate the coefficient matrix $\mathfrak{A}(z)$ to the Weyl matrix appearing in the FMI, \cref{eq:FMI_Weyl}.
Roughly speaking, the idea is to split apart the $z$-dependence of the Weyl matrix symmetrically in such a way that one can read off the form of the coefficient matrix $\mathfrak{A}(z)$.
The concrete goal is to factorize the Weyl matrix into the form
\begin{align}
    W(z^*,z) = \mathfrak{A}^{\dag-1}(z) J_2 \mathfrak{A}^{-1}(z), \label{eq:Weyl_factorization_goal}
\end{align}
where $\mathfrak{A}(z)$ is a matrix-valued analytic function in the upper half-plane.
When this form has been achieved, the \emph{Ansatz} reduces the FMI in the upper half-plane to \cref{eq:pq_J2_positive}, which is positive by assumption.
The key role is played by the ``splitting matrix''\footnote{Perhaps surprisingly, the matrix-valued function $T$ is not given a name in Kovalishina's work \cite{Kovalishina1984}, despite its importance to the argument. Consequently, the name ``splitting matrix'' is novel to this work.} $T(z_1, z_2)$
\begin{align}
    T(z_1, z_2) \equiv I + \mathscr{H}(z_1, z_2) J_2,
\end{align}
which enjoys a useful transitivity property
\begin{align}
    T(z_1, z_2) T(z_2, z_3) = T(z_1, z_3).
\end{align}
The splitting matrix is related to the Weyl matrix by 
\begin{align}
    W(z_1, z_2) = J_2 T(z_1, z_2).
\end{align}
Given an arbitrary ``support point" $z_0$ in the upper half-plane, one can show that~\cite{Kovalishina1984}
\begin{align}
    W(z^*,z) = T^{\dag-1}(z, z_0) W(z^*_0,z_0) T^{-1}(z,z_0),
\end{align}
which reduces the problem to that of factoring the constant matrix $W(z^*_0, z_0)$.
This factorization can be carried out for  generic $z_0$, although a judicious choice simplifies subsequent algebraic manipulations.
Without loss of generality, Kovalishina has shown that one may use the boundary point $z_0=0$.\footnote{In applications, it may be interesting to explore possible numerical advantages to other choices for the support point $z_0$.}
This choice gives
\begin{align}
    \left. W(z^*_0, z_0)\right|_{z_0=0} = J_2,
\end{align}
from which the FMI in \cref{eq:FMI_Weyl} then becomes
\begin{align}
    \left[ G^\dagger(z), I\right] \frac{T^{\dag-1}(z, z_0) J_2 T^{-1}(z,z_0)}{(z-z^*)/i}
    \begin{bmatrix}
        G(z) \\ I
    \end{bmatrix} \succeq 0.
\end{align}
But the coefficient matrix for the solution can then be read off immediately by comparison to \cref{eq:Weyl_factorization_goal}:
\begin{align}
\mathfrak{A}(z)
&= T(z,z_0)\\
&= I - i 
\times\begin{bmatrix}
    -\bm{c}^\dagger(z^*) \\
    \bm{b}^\dag(z^*)
\end{bmatrix}
\hankel^{-1} 
\left[ -\bm{c}(0), \bm{b}(0) \right].
\end{align}
The technical part of Kovalishina's proof consists of showing, with due care for analyticity and the existence of all the required inverses,
that the inclusion goes both ways, i.e., 
that the \emph{Ansatz} satisfies the FMI \emph{and} and any solution of the FMI can be written in the from of the \emph{Ansatz}.
Kovalishina has further shown that the coefficient matrix $\mathfrak{A}(z)$ is $J_2$-expanding in the upper half-plane and $J_2$-unitary on the real line, i.e.,
\begin{align}
    \mathfrak{A}^\dag(z) J_2 \mathfrak{A}(z) - J_2 \succeq 0 && z \in \C^+ \\
    \mathfrak{A}^\dag(z) J_2 \mathfrak{A}(z) - J_2 = 0 && z \in \R,\phantom{.} \label{eq:A_J_unitary}
\end{align}
and emphasized that these properties of the solution make manifest the underlying $J$-theoretic structure of the problem.
One immediate consequence of \cref{eq:A_J_unitary} is that the Weyl matrix $W(z^*, z)$ is equal to $J_2$ for $z \in \R$, which implies that the only condition placed on $G(z)$ for $z$ limiting towards the real axis is \cref{eq:stieltjes_positvity}. This statement is analogous to the similar statement in Ref.~\cite{Bergamaschi:2023xzx} that the Wertevorrat grows to encapsulates the full upper half plane whenever the target smearing approaches zero.

\subsection{Proof of theorem on the Weyl matrix ball\label{sec:Weyl_ball}}

\begin{proof}[Sketch of proof for \cref{thm:weyl_matrix_ball}]
Suppose that $G$ satisfies the positivity condition
\begin{align*}
    (G^\dag, I) W \begin{bmatrix}
        G \\ I
    \end{bmatrix} \succeq 0.
\end{align*}    
Since the coefficient matrix $\mathfrak{A}$ is $J_2$-expanding in the upper half-plane, the Weyl matrix $W=\mathfrak{A}^{\dagger-1} J_2 \mathfrak{A}^{-1}$  is $J_2$-contractive in the upper half-plane and may therefore parameterized as
\begin{align}
    W = \begin{bmatrix} -R & S \\ S^\dagger & -T \end{bmatrix},
\end{align}
with $R\succ0$ and $S^\dagger R^{-1} S -T \succ 0$~\cite{Kovalishina1984} using \cref{lemma:block_matrices}.\footnote{On the real line, the fact that $W=J_2$ implies that $R=T=0$, so formulas involving $R^{-1}$ should be understood in light of the discussion following \cref{eq:A_J_unitary}.}
With this parameterization, the positivity condition becomes
\begin{align}
    -G^\dag R G + G^\dag S + S^\dag G - T  \succeq 0.
\end{align}
Factoring the first three terms yields
\begin{align}
\begin{split}
&\left(G^\dag R^{1/2} - S^\dagger R^{-1/2}\right)
\left( R^{1/2} G - R^{-1/2} S\right) \\
&\quad\preceq 
\left( S^\dag R^{-1} S - T \right)^{1/2}
\left( S^\dag R^{-1} S - T \right)^{1/2}.
\end{split}
\end{align}
But since the first term is smaller (or equal) to the second term, there exists a contraction $X$ ($X^\dag X \preceq I$) that will saturate the bound, namely,
\begin{align}
    R^{1/2} G - R^{-1/2} S
    = X \left( S^\dag R^{-1/2} S - T \right).
\end{align}
Solving for $G$ gives
\begin{align}
    G = R^{-1} S + R^{-1/2} X \left(S^\dagger R^{-1} S-T\right)^{1/2},
\end{align}
which shows that $G$ fills the Weyl matrix ball, as claimed.
\end{proof}

\subsection{Proof of projection lemma\label{sec:projection_lemma}}

\begin{proof}[Proof of \cref{lemma:weyl-ball-project}]
First, the left hand side of \cref{eq:weyl-ball-proj} can be simplified as
\begin{align}
\begin{split}
\vecl^\dag &\ball(C; A, B) \vecr\\
    &= \{\vecl^\dag C \vecr + \vecl^\dag AXB \vecr\mid I - X^\dag X \geq 0 \} \\
    &= \{\vecl^\dag C \vecr + (A^\dag \vecl)^\dag X (B\vecr) \mid I - X^\dag X \geq 0 \}
\end{split}
\end{align}
Shifting by a constant, it suffices to consider the case $C = 0$, i.e., it suffices to prove the equality
\begin{equation}
    \{(A^\dag \vecl)^\dag X (B\vecr) \mid I - X^\dag X \geq 0 \} = \disk(0, r).
\end{equation}
This equality can be proven by considering both inclusions separately:

Consider first the inclusion of the projected ball within this disk.
% $\subseteq$: 
Let $z \in \vecl^\dag \ball(0; A, B) \vecr$, so $z = (A^\dag \vecl) X (B\vecr)$ with $X$ contractive.
Since $X$ is contractive, one may compute
\begin{equation}
\begin{aligned}
|z| &= \abs{(A^\dag \vecl) X (B\vecr)} \\
&\leq \norm{A^\dag \vecl} \norm{B\vecr} \norm{X} \\
&\leq \norm{A^\dag \vecl} \norm{B\vecr} \\
&= r
\end{aligned}
\end{equation}
and hence $z \in \disk(0, r)$ as expected.

Consider next the inclusion of the disk within the projected ball.
% $\supseteq$: 
Let $z \in \disk(0, r)$, take $X = z (A^\dag \vecl) (B\vecr)^\dag/r^2$.
Then $X$ is contractive since
\begin{equation}
    \norm{X} = \abs{z} \frac{\norm{A^\dag \vecl} \norm{B \vecr}}{r^2}
    = \frac{\abs{z}}{r} \leq 1
\end{equation}
and furthermore
\begin{equation}
(A^\dag \vecl)^\dag X (B\vecr) = \frac{z}{r^2} \norm{A^\dag \vecl}^2 \norm{B\vecr}^2
= z
\end{equation}
and hence $z \in \vecl^\dag \ball(0; A, B) \vecr$ as expected.
\end{proof}

\clearpage
\bibliography{main.bib}

\end{document}